**Bradley-Sardin telecentric telescope for enhanced detection of the aberration of stellar light**

*G. Sardin*
*Applied Physics Department, University of Barcelona - Barcelona, Spain*

**Abstract**

A telescope specifically designed for the observation of the stellar aberration of light is proposed. It is distinctive in two main features: a lengthy collimated beam and an adjustable position of the photo-detector along the telescope axis, so the beam length-of-flight can be varied from zero to the telescope total length. This is achieved by means of a telecentric objective projecting the collimated beam up to a movable CCD camera. The spot position on a high-resolution photo-detector array is recorded, and the data are transferred to a computer and treated by a beam analyser software. The telescope aims to measure with high accuracy the aberration of stellar light due to the earth orbital motion. An alternative would consist of fixing the telecentric objective at the top of a cliff by means of an anti-vibration clamp with a high-precision directional mount. The detector would be fixed on a micrometric positioner, placed on the foot of the cliff at the beam position of fall. This way, the beam time-of-flight can be considerably increased, and hence the subsequent spot shift. For a beam length of 300 m the resulting shift would be of 30 mm, and so in a year the spot would describe a circle of 60 mm diameter. Optionally, a diode laser may be fixed at the centre of the telescope objective in order to contrast the behaviour of local and stellar light.

**I. Introduction**

Let us at first briefly acquaint the Bradley effect. The aberration of stellar light was first interpreted by Bradley in 1726 (1). He observed that the position of stars traces out yearly small ellipses and he related this effect to the earth speed around the sun. Since all ellipses from stars in a close direction had the same size they thus could not be due to the parallax effect, which would imply all stars to be equally distant from the earth. Furthermore, at two positions of the earth, six months apart, the speed vectors have opposite directions and consequently the two ensuing aberration angles are opposite. From these facts Bradley concluded that the effect observed was due to the aberration of light uniquely induced by the observer motion. It can be simply stated as:

tn $\varphi$ = v / c

Stellar aberration has been proved to be independent of the source speed and to have just a local origin exclusively based on the earth orbital speed, being of 30 km/s relative to the sun (2-9). The corresponding aberration angle $\varphi$ is equal to 20.5 arc-s, value called the constant of stellar aberration.

**II. a. Specificity of the telescope**

This telescope differs from standard telescopes in that its finality is not to enlarge the image of stars but at the contrary to reduce it to a point on the array detector and to track any displacement of its position on it. Optically it differs in that its objective is telecentric (10), so incident stellar light rays are not straightforwardly focused but stay parallel and emerge as a narrower collimated beam. The ensuing advantage consists in allowing to position the ocular lens at any desired distance from the objective and hence to elect at will the beam time-of-flight within the telescope. The punctual image of the star is collected by means of a high resolution CCD camera. The spot temporal position on the array detector is transferred to a computer and analysed by a beam profiler software (11).

Since the camera is mobile all along the beam path up to the objective the beam time-of-flight can be varied from zero to a maximum value $t_o$ and so the beam-spot shift $\Delta x$ caused by the concurrent drift of the conveying inertial system varies from zero to a maximum value fixed by the telescope length. The spot shift is proportional to the speed of the inertial system and to the beam time-of-flight from the objective to the photo-detector. So, for a given inertial speed the longer the distance objective-detector the larger the spot shift.

**II. b. Measure of stellar aberration**

For a beam path of 3 m length (d) and the earth speed (v) around the sun being of about 30 km (its average speed is 29.79 km), the stellar spot will describe in a year a circle with radius of about 300 microns. In effect, the time (t) that light lasts (time-of-flight) to cover 3 m is:

$t = d / c = 3 / 3 * 10^8 = 10^{-8}$ s

The aberration shift ($\Delta x$), induced by the meanwhile earth lateral drift is:

$\Delta x = v * t = (3*10^4)*10^{-8} = 3*10^{-4}$ m = 300 $\mu$m

However, this shift is not straightforwardly observable since there is no referential origin. So, it requires a long time to become discernible, i.e. until the trace of the beam spot on the array detector gets extended enough as to allow outlining a sector of the circle it describes in a year and so to extract its radius. A time of about 3 months may be necessary to get a reliable value.

Since the telecentric objective allows lengthening the beam path at will an alternative would consist of fixing it e.g. at the top of a disused chimney or at the terrace border of a sturdy building free of local vibrations and the detector at a distant underneath window. Still better, the telecentric objective could be fixed at the top of a cliff through a high precision adjustable clamp. The detector would be mounted on micrometric positioner, placed at the bottom of the cliff, right at the beam position of fall. This way, the beam time-of-flight can be considerably increased, and hence the subsequent spot shift. For a beam length of 300 m the resulting shift would be 100 times 300 $\mu$m, i.e. 30 mm. So, in a year the spot would describe a circle of 60 mm diameter.

**II. c. Laser option**

A pointing diode laser may be added to the telescope, fixed at the centre of its objective lens, allowing an extra assessment of the Bradley effect. The spot from the star and the laser are at first adjusted so that they superpose. After while the time dependence of the position on the array detector of the spot from the star and from the laser is simultaneously tracked. In the course of one year the apparent position of the star will describe a quasi-circular ellipse, but since no aberration has been detected from any local source the spot from the laser beam placed on the telescope objective will remain still on the array-detector at its initial position (7).

**III. Conclusion**

The essential advantage of the telescope proposed resides in that it allows an enhanced gauging of Bradley aberration and thus of the earth orbital speed, through its lengthy collimated beam. Also, it is adequate for hobbyists since relatively easy to assemble and so to observe the yearly Bradley aberration of light.

Figure 1: Telecentric telescope

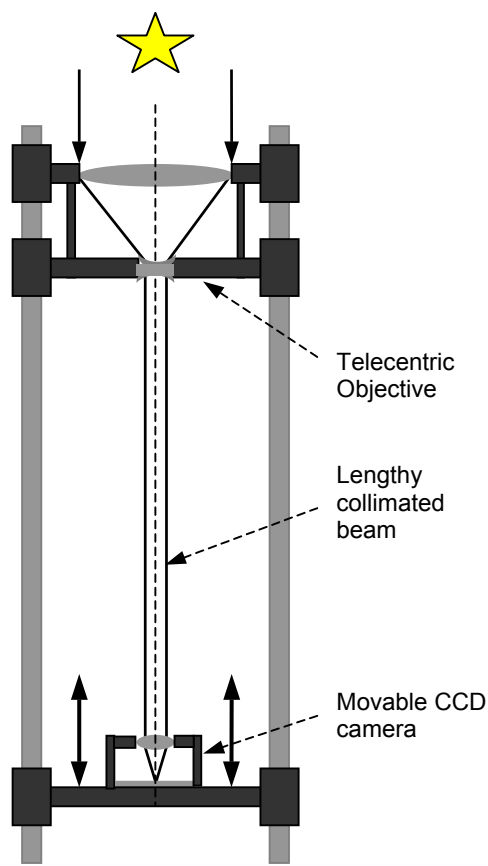